\documentclass{article}

\usepackage[preprint]{neurips_2025}

\usepackage[utf8]{inputenc} 
\usepackage[T1]{fontenc}    
\usepackage{hyperref}       
\usepackage{url}            
\usepackage{booktabs}       
\usepackage{amsfonts}       
\usepackage{nicefrac}       
\usepackage{microtype}      
\usepackage{xcolor}         
\usepackage{graphicx}
\usepackage{amsmath}
\usepackage{float}  
\usepackage{placeins}  
\usepackage{verbatim}

\usepackage{orcidlink} 

\usepackage{booktabs}
\usepackage{array}    
\usepackage{makecell} 

\usepackage{titlesec}
\titlespacing*{\section}
  {0pt}      
  {5.5pt}     
  {5.5pt}    

\titlespacing*{\subsection}
  {0pt}      
  {5.5pt}     
  {5.5pt}    

\titlespacing*{\subsubsection}
  {0pt}      
  {5.5pt}     
  {5.5pt}    

\usepackage{natbib}

\setcitestyle{numbers,square}

\title{EEGReXferNet: A Lightweight Gen-AI Framework for EEG Subspace Reconstruction via Cross-Subject Transfer Learning and Channel-Aware Embedding}

\author{
Shantanu Sarkar\textsuperscript{1*}\orcidlink{0000-0002-5125-6925},
Piotr Nabrzyski\textsuperscript{1,2}, 
Saurabh Prasad\textsuperscript{1}\orcidlink{0000-0003-3729-9360}, 
Jose Luis Contreras-Vidal\textsuperscript{1}\orcidlink{0000-0002-6499-1208} \\
\\
\textsuperscript{1} IUCRC BRAIN Center, Cullen College of Engineering, University of Houston, Houston, TX, USA \\
\textsuperscript{2} Department of Computer Engineering, Purdue University, West Lafayette, IN, USA \\
\textbf{*Corresponding author:} Shantanu Sarkar (\href{mailto:shantanu75@gmail.com}{shantanu75@gmail.com})
}

\begin{document}

\maketitle

\begin{abstract}
Electroencephalography (EEG) is a widely used non-invasive technique for monitoring brain activity, but low signal-to-noise ratios (SNR) due to various artifacts often compromise its utility. Conventional artifact removal methods require manual intervention or risk suppressing critical neural features during filtering/reconstruction. Recent advances in generative models, including Variational Autoencoders (VAEs) and Generative Adversarial Networks (GANs), have shown promise for EEG reconstruction; however, these approaches often lack integrated temporal-spectral-spatial sensitivity and are computationally intensive, limiting their suitability for real-time applications like brain–computer interfaces (BCIs). To overcome these challenges, we introduce \textbf{\textit{EEGReXferNet}}, a lightweight Gen-AI framework for EEG subspace reconstruction via cross-subject transfer learning - developed using Keras TensorFlow (v2.15.1). \textbf{\textit{EEGReXferNet}} employs a modular architecture that leverages volume conduction across neighboring channels, band-specific convolution encoding, and dynamic latent feature extraction through sliding windows. By integrating reference-based scaling, the framework ensures continuity across successive windows and generalizes effectively across subjects. This design improves spatial-temporal-spectral resolution (mean PSD correlation $\geq$ 0.95; mean spectrogram RV-Coefficient $\geq$ 0.85), reduces total weights by $\sim45\%$ to mitigate overfitting, and maintains computational efficiency for robust, real-time EEG preprocessing in neurophysiological and BCI applications.
\end{abstract}

\section{Introduction}
Electroencephalography (EEG) is a non-invasive method that records brain activity via scalp electrodes by amplifying spontaneous potentials, providing insight into spatial, temporal, and spectral patterns associated with sensory and cognitive processes \cite{Cohen2017, Siuly2016}. EEG is widely used in clinical and neuroscience, cognitive and psychiatric studies due to its non-invasive, safe, and relatively inexpensive nature, its ability to directly measure neuronal electrical activity, and its real-time capability, which has also made it increasingly popular in brain-computer interface (BCI) applications\cite{Craik2019, Craik2023, Sugden2023}.
\\
EEG are highly susceptible to artifacts from physiological (e.g., ocular, muscular) and non-physiological (e.g., powerline, motion) sources, resulting in low signal-to-noise ratios (SNR) with reduced interpretability\cite{Craik2023, Tandle2016, Uriguen2015, Jiang2019, Rashmi2022}. While denoising EEG signals by removing components from various unwanted sources, precise filtering is essential to filter artifacts that share frequency content similar to the underlying EEG features \cite{Kilicarslan2021}. For closed-loop implementation of the BCI system, the event-locked nature of many types of artifacts complicate real-time neural decoding \cite{Kilicarslan2021}. 
\\
Substantial efforts have been devoted to developing algorithms for EEG artifact removal, each offering distinct advantages and limitations \cite{Craik2023, Uriguen2015, Jiang2019, Rashmi2022}. Independent Component Analysis (ICA) utilizing blind source separation (BSS) techniques is the most widely used methods; however, it requires manual intervention to exclude artifacts \cite{Jiang2019, Rashmi2022}. To automate artifact removal, early approaches employed channel-wise statistical thresholding to eliminate abnormal activity patterns \cite{Chang2020} - one such method is wavelet-based filtering with sparsity constraints \cite{Geetha2011}; however, these techniques often compromise the full feature representation of the affected channel or suppress spectral features in sparsity-enforced bands. Adaptive filtering techniques, such as H-Infinity \cite{Kilicarslan2016, Kilicarslan2019}, have demonstrated promising results in real-time artifact denoising, but the need for a reference noise signal limits their application. The BSS methods like ICA and Principal Component Analysis (PCA) estimate the sources of artifacts considering a mixing matrix for original and observed signals by decomposing the EEG into transformed spaces\cite{Uriguen2015, Jiang2019, Rashmi2022}. The Artifact Subspace Reconstruction (ASR) decomposes EEG into a principal component (PC) space derived from clean data covariance and removes artifacts by suppressing components with abnormally high variance\cite{Chang2020}. Transformed spaces like PC space are linear combinations of all channels; suppressing contaminated components in transformed space may alter the data structure, potentially losing essential features \cite{Oladipupo2014}.
\\
With recent developments in Generative AI, the Variational Autoencoder (VAE)- and Generative Adversarial Network (GAN)-based models are gaining popularity as an EEG pre-processing method \cite{Hwaidi2021,Sun2023}. Generative reconstruction methods often overlook spatial channel relationships, use heavy encoder-decoder models with weak temporal-spectral coupling, and lack consistent mapping across successive sliding windows. To address these limitations, we propose \textbf{\textit{EEGReXferNet}} - a novel, lower memory footprint architecture for EEG channel subspace reconstruction that leverages volume conduction across neighboring channels\cite{Cohen2017, Lagerlund2016}, integrates spatial structure and dynamic temporal-spectral encoding, and applies reference scaling ($\mu_{Ref},\ \sigma_{Ref}$) for temporal continuity. \textbf{\textit{EEGReXferNet}} is trained via Cross-Subject Transfer Learning to enhance generalizability across individuals. 
\section{Model Architecture}
\begin{figure}[b!]
    \centering
    \includegraphics[width=\linewidth]{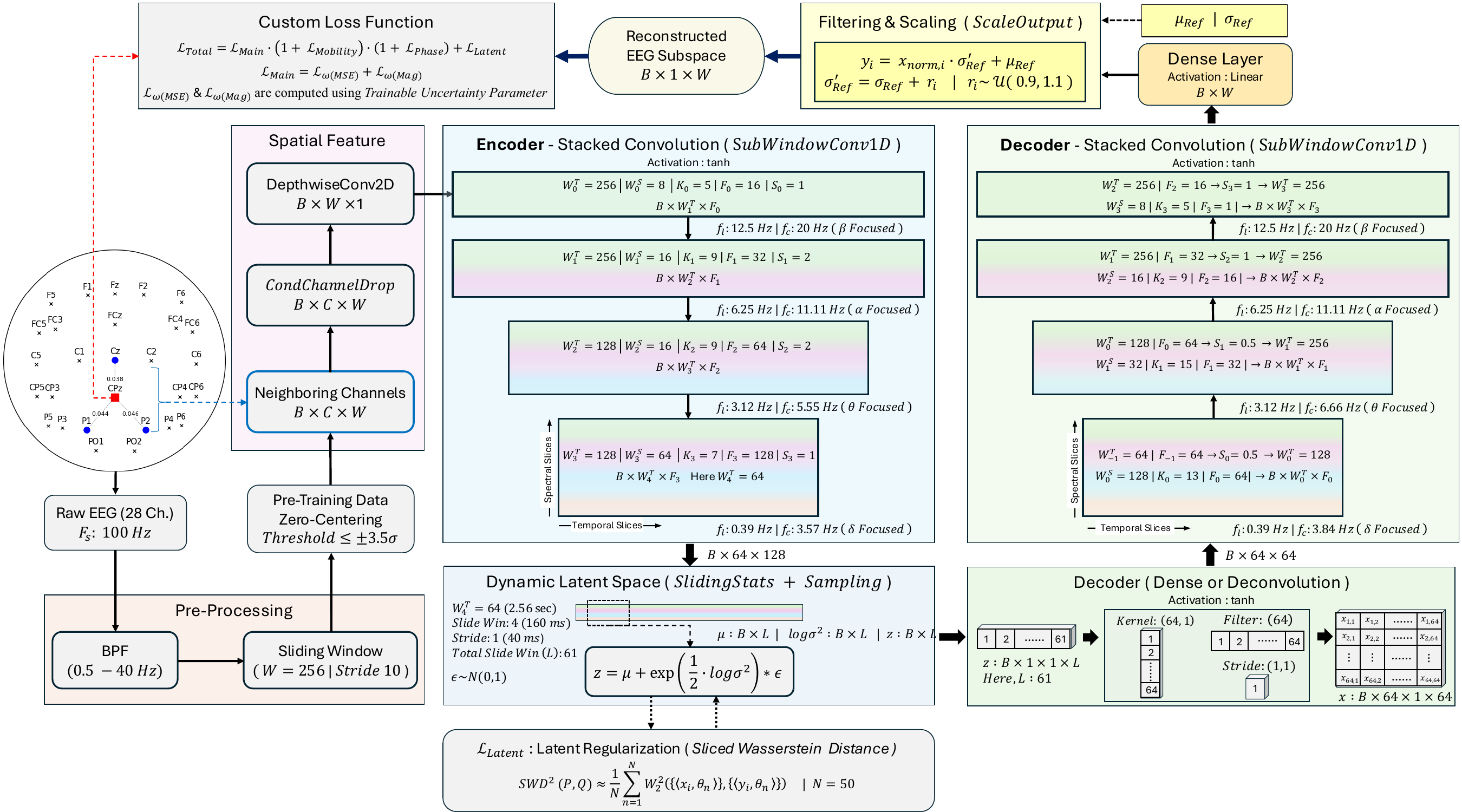} 
    \caption{Overview of \textit{EEGReXferNet}  architecture illustrating key processing blocks and workflow.}
    \label{fig:Fig1}
\end{figure}

\textbf{\textit{EEGReXferNet}} adopts a modular design with task-specific layers and a custom loss function, as shown in Figure \ref{fig:Fig1}. Implemented in Keras TensorFlow (v2.15.1), it supports scalable, interpretable, GPU-efficient modeling. Each layer and loss function are detailed below.

\textbf{Neighborhood-Driven Input Selection:} The model takes multichannel EEG windows (B, C, W) as input, from which it selects selects neighboring channels via a predefined dictionary. This dictionary maps each EEG channel to the indices of its nearest neighbors (L2 distance < 0.05) in the 10–20 system \cite{Bcker1994}. During training, conditional channel dropout is applied using \texttt{SpatialDropout1D}, which drops one or two neighbors based on channel counts ($\leq 3$). Finally, depth-wise convolution aggregates spatial dependencies into (B, 1, W). Here, B is the batch size, C is the total number of channels, and W is the size of the sliding window.
\\
\textbf{Sub-Window Convolution Encoding Block:} The output of the “Neighborhood-Driven Input Selection” is passed through a series of custom \texttt{SubWindowConv1D} layers. Hartmann et al. showed that stacked convolutions extract fine-grained spectral features from EEG signals, with each layer specializing in distinct spectral characteristics \cite{Hartmann2018}. Building on this insight, we implemented band-specific stacked convolutions using a custom-designed \texttt{SubWindowConv1D} layer, parameterized by kernel size, stride, filters, sub-window size, and tanh activation. To enforce spectral selectivity, we tailored the kernel sizes, strides, and sub-window configurations for each iteration. Parameters and targeted frequency bands are shown in Figure \ref{fig:Fig1}.
\\
\textbf{Sliding Stats Layer:} This layer applies a sliding window mechanism to segment the input into overlapping temporal frames, where two lightweight dense layers estimate latent statistics. It enhances temporal resolution while reducing parameters by 45\% compared to a dense layer for a 32-D latent space, thereby lowering memory use and mitigating overfitting—making it well suited for real-time, low-SNR applications. We used 160-ms sliding windows with a 40-ms stride to capture fine-grained, microstate-level temporal dynamics \cite{Michel2018}.
\\
\textbf{Sampling Layer and Latent Regularization:} The \texttt{SamplingLyr} performs reparametrized sampling following the classical VAE framework by injecting Gaussian noise, enabling differentiable transformations of encoder outputs \cite{Kingma2013}. For latent regularization, we replace KL-divergence (KLD) with the geometry-aware, sample-based Sliced Wasserstein Distance (SWD)\cite{Kolouri2016,Kolouri2018} using 50 projections, improving gradient stability and reducing min–max conflicts in high-dimensional latent spaces.
\\
\textbf{Base Decoding Using Transposed Convolution:} This block uses transposed convolution (deconvolution) to progressively up-sample the latent vector z into a structured feature map. Unlike dense layers, it provides spatially structured, parameter-efficient up-sampling\cite{Zeiler2010}.
\\
\textbf{Sub-Window Convolution Decoding Block:}
The output from the “Base Decoding” is passed through a series of custom \texttt{SubWindowConv1D} layers. Like encoding, each layer is parameterized by tailored kernel size, stride, number of filters, and sub-window size to implement band-specific stacked convolutions. The selected parameters for each iteration are shown in Figure \ref{fig:Fig1} for the respective targeted frequency bands.
\\
\textbf{Linear Activation Layer:} The output from the “Sub-Window Convolution Decoding Block” is flattened and passed to a linear-dense layer to produce the continuous-valued reconstruction.
\\
\textbf{Filtering and Scaling Layer:} After decoding, the reconstructed outputs are passed sequentially through the \texttt{RemoveOutlier}  and \texttt{ScaleOutput} Layer layers to produce a sample-wise adaptive reconstruction. The \texttt{RemoveOutlier} layer clips values beyond a z-score threshold, mitigating any spurious deviations in the decoded signal. Subsequently, the \texttt{ScaleOutput} Layer normalizes by sample-wise mean and SD and then rescales it using reference statistics ($\mu_{Ref},\ \sigma_{Ref}$). During training, reference statistics are computed from clean EEG segments, whereas for reconstruction of noisy subspaces, the layer utilizes statistics from previously clean segments. Random scaling factors ($\pm10\%$) were applied to achieve stochastic regularization.
\\
\textbf{Loss Function:} We used a latent regularization loss ($\mathcal{L}_{Latent}$) via SWD \cite{Kolouri2016,Kolouri2018}, combined with a custom loss aggregating domain-specific Mean Square Error (MSE) and Hjorth mobility\cite{Mehmood2017}. Temporal-MSE ($\mathcal{L}_{mse}^\omega$) and Magnitude-Spectrum-MSE ($\mathcal{L}_{mag}^\omega$) are weighted by a trainable uncertainty parameter\cite{Cipolla2018} and multiplicatively coupled\cite{Coleman2025} with Phase-Spectrum-MSE ($\mathcal{L}_{phase}$) and Hjorth mobility loss ($\mathcal{L}_{mobilty}$). The total loss function is defined by Equ. (\ref{eq:equ1}).
\begin{equation}\label{eq:equ1}
\mathcal{L}_{\text{Total}} = 
\big(\mathcal{L}_{\text{mse}}^{\omega} + \mathcal{L}_{\text{mag}}^{\omega}\big)
\cdot \big(\mathcal{L}_{\text{mobility}} + 1\big)
\cdot \big(\mathcal{L}_{\text{phase}} + 1\big)
+ \mathcal{L}_{\text{latent}}
\end{equation}
\FloatBarrier  
\section{Materials and Methods}
\subsection{Dataset }
Our goal is to reconstruct contaminated EEG subspaces in real time and improve motor imagery (MI) decoding, focusing on BCI Competition IV (Dataset 1) \cite{Blankertz2007}. The dataset includes 59-channel EEG from 7 subjects (a, b, g, f: human; others synthetic) performing binary MI tasks. Data were band-pass filtered (BPF, 0.05–200 Hz) and digitized at 1000/100 Hz. This study used 100 Hz calibration data from human subjects, selecting 28 channels (Figure \ref{fig:Fig1}) matching our paradigm.
\subsection{Pre-processing}
To further preprocess the EEG, we applied a $6^{th}$-order zero-phase Butterworth BPF (0.5 - 40 Hz). Following BPF, to create the EEG subspace, we used a sliding window of 2.56 seconds to ensure the minimum frequency $\geq\ 0.4 Hz$, with a stride of 100 ms. Within each sliding window, the channel-wise signal was re-centered to align with the global channel mean and added with a small random perturbation ($\pm10\%$). We stratified the sliding windows into \textbf{‘Clean’} and \textbf{‘Noisy’} categories based on amplitude thresholding. Under \textbf{‘Clean’}, we considered only those EEG windows, where the channel-wise amplitude range fell within specified limits ($>\pm3.5\sigma$) and were utilized to train the model. The remaining windows were categorized as \textbf{‘Noisy’} and used for evaluation.
\subsection{Methodology}
\begin{figure}[b]
    \centering
    \includegraphics[width=\linewidth]{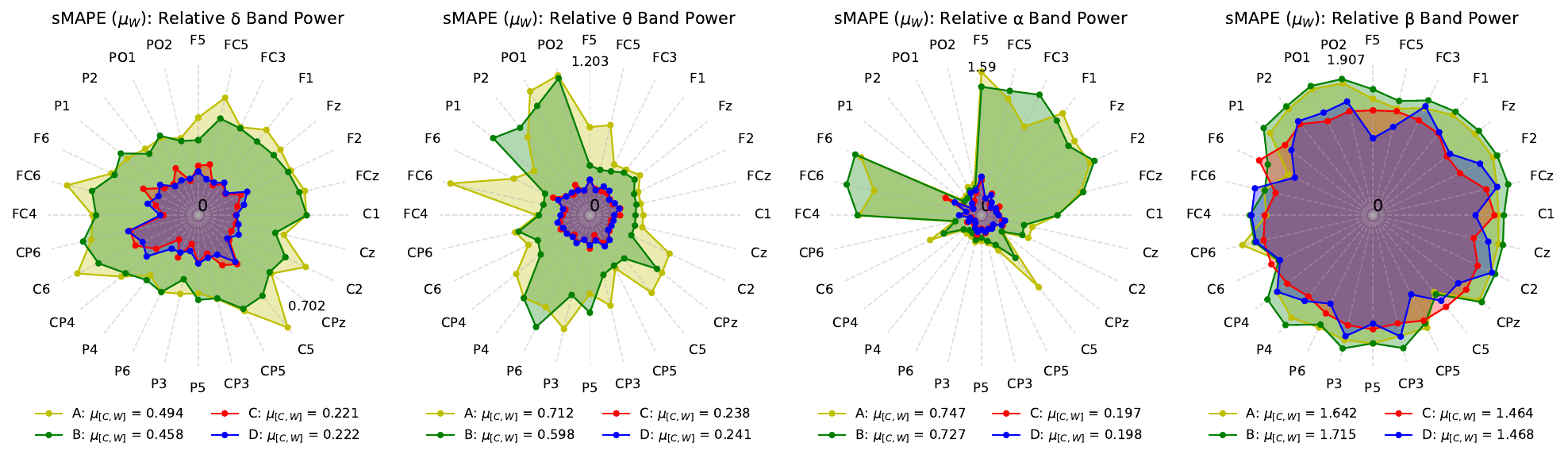}
    \caption{Subject ‘a’: sMAPE across EEG channels/windows for relative $\delta$, $\theta$, $\alpha$, and $\beta$ band power.}
    \label{fig:Fig2}
\end{figure}

To evaluate the key components of \textbf{\textit{EEGReXferNet}}, we performed an ablation study across four configurations: (i) KLD (Model A) vs. SWD (Model B), (ii) fixed (Models A \& B) vs. dynamic latent space (Models C \& D), and (iii) dense (Model C) vs. deconvolution decoding (Model D).
\\
Models were trained on clean EEG data from three subjects, leaving one out for evaluation. Clean segment reconstruction used window/channel-wise scaling based on their stats ($\mu, \sigma$). For reconstruction error metrics, we followed FDA work on EEG feature consistency \cite{Nahmias2019,Nahmias2021} and used Symmetric Mean Absolute Percentage Error (sMAPE) across channels/windows for relative $\delta$, $\theta$, $\alpha$, $\beta$ power (Figure \ref{fig:Fig2}), temporal/spectral entropy, and mobility, along with JS-Divergence. EEG channel-wise probability densities were estimated using the Dual Polynomial Regression method from the \texttt{estimatePDF} Python package\cite{Sarkar2025_estimatePDF}. Furthermore, we evaluated the MSE across the time, frequency (phase and magnitude), and time–frequency (spectrogram) domains. Noisy-window reconstruction utilized reference statistics ($\mu_{\text{Ref}}, \sigma_{\text{Ref}}$) from preceding clean windows, and was evaluated using PSD correlation (Pearson) and spectrogram similarity (RV coefficient). Model performance was compared using metrics across all channels with Friedman tests and post-hoc Nemenyi and Wilcoxon tests ($\alpha=0.01$). For downstream classification, we trained \textbf{\textit{EEGNet-8-2}}\cite{Lawhern2018} with clean EEG windows and tested noisy ones as the baseline; misclassified windows were then reconstructed with Model D and re-evaluated.
\\
All models were trained with a batch size of 64 using Adam optimizer (Learning rate=0.001), early stopping (patience=25), up to 250 epochs, with 20\% validation. For reproducibility, the backend was reset with a fixed seed prior to training for each channel, and a seed was similarly used for downstream evaluation. To train \textbf{\textit{EEGReXferNet}} used Sabine Cluster (1 node, 28 cores, 1 GPU [16 GB]) and the rest of the task was done with a MacBook Pro (Apple M1 Max, 10 cores, 1 GPU).
\section{Results}
\begin{figure}[t]
    \centering
    \includegraphics[width=\linewidth]{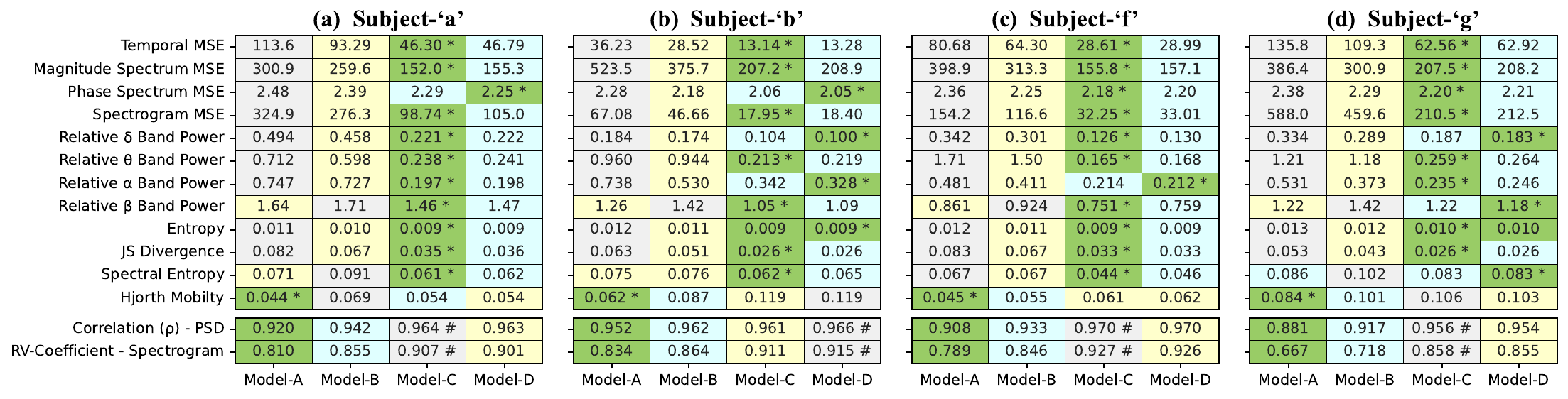}
    \caption{Model comparison across subjects using EEG metrics. Heatmaps show Wilcoxon ranks for (top) clean and (bottom) noisy data. In (top), * marks the best (min), in (bottom), \# marks the best (max). Cells show mean scores. Color scale: green (min) $\rightarrow$ cyan $\rightarrow$ yellow $\rightarrow$ gray (max).}
    \label{fig:Fig3}
\end{figure}

\begin{figure}[H]
    \centering
    \includegraphics[width=\linewidth]{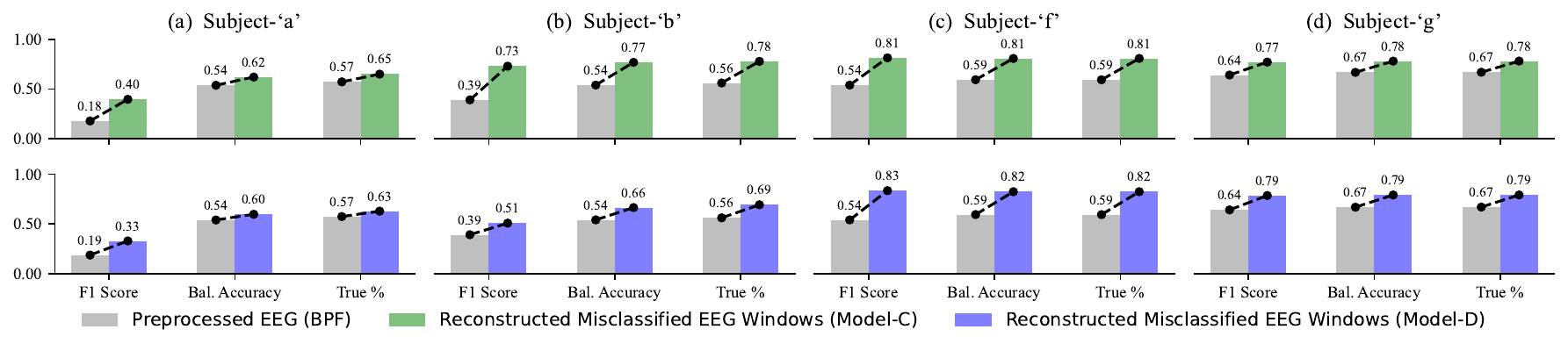}
    \caption{Comparison of accuracy metrics (Downstream classification using EEGNet-8-2) across subjects - Baseline vs. Reconstructed misclassified EEG windows via Model-C and D.}
    \label{fig:Fig4}
\end{figure}

Friedman tests with post-hoc Nemenyi indicated that metrics differed significantly across models in most cases, which was further supported by Wilcoxon pairwise tests. Figure \ref{fig:Fig3} presents the Wilcoxon rank-based heatmaps with overall mean scores. Models C and D, with dynamic latent space (45\% fewer weights), consistently outperformed Models A and B across metrics, except for Hjorth mobility, where differences were minimal ($\leq0.05$). Between the latent regularization strategies, SWD outperformed KLD, yielding more consistent improvements across subjects. 
Subject- and metric-specific differences were observed between Model C (dense) and Model D (deconvolution); however, their overall performance remained statistically consistent. Correlation analyses on noisy windows mirrored the clean-window findings: Models C and D exhibited higher correlations than Models A and B for both PSD and spectrogram-based measures. Downstream classification showed a marked improvement in accuracy metrics across all subjects when previously misclassified noisy EEG windows were reconstructed using Model-C and Model-D, and subsequently re-evaluated via EEGNet-8-2, as illustrated in Figure~\ref{fig:Fig4}. Model-C yielded superior performance for Subjects-`a' and `b', while Model-D outperformed for Subjects-`f' and `g'. Model-D exhibited the lowest mean training time across EEG channels ($\sim 11\ \text{min}$), in contrast to Model-A, which showed the highest ($\sim 16\ \text{min}$). Mean inference times per sliding window across EEG channels showed minimal variation: Model-A (0.75 ms), Model-B (0.77 ms), Model-C (0.78 ms), and Model-D (0.78 ms).


\section{Conclusion}
This work introduced \textbf{\textit{EEGReXferNet}}, a novel lightweight Gen-AI framework for EEG subspace reconstruction that integrates temporal, spectral, and spatial sensitivity while maintaining computational efficiency. By leveraging cross-subject transfer learning and reference-based scaling, the model demonstrated robust generalization and continuity across successive windows. Models C and D, leveraging dynamic latent spaces with SWD regularization, consistently outperformed baselines across all metrics, except Hjorth mobility (minimal difference). For Model-C and D, although performance varied across subjects and metrics, both models exhibited consistent statistical behavior across the full evaluation spectrum. Furthermore, \textbf{\textit{EEGReXferNet}} maintained consistently low inference latency (
$< 1\ \text{ms}$), underscoring its suitability for real-time neurophysiological applications. Future work should evaluate the model on larger and diverse datasets (beyond MI), integrate adaptive artifact detection for real-time use, and analyze inner features to understand learned representations.
\paragraph{Code Availability: }
The code is available on GitHub (\href{https://github.com/ShanSarkar75/EEGReXferNet}{\textbf{\textit{ShanSarkar75/EEGReXferNet}}}), and the reconstructed EEG window data (.npz) are shared via FigShare (DOI: \href{https://doi.org/10.6084/m9.figshare.30343642}{10.6084/m9.figshare.30343642}).
\begin{ack}
Supported by NSF IUCRC BRAIN and the NSF REU Site on Regulatory Science (Award \#2349657). We thank Ramin Bighamian (FDA) and Xiaoqian Jiang (UTHealth) for their insightful feedback
\end{ack}

\section*{References}

\medskip

{
\small

\bibliographystyle{unsrtnat}
\bibliography{References}

}

\newpage
\appendix

\section{Technical Appendices and Supplementary Material}

\subsection{Additional Tables}
\begin{table}[ht]
  \caption{Comparison of model configurations used for ablation study.}
  \label{tab:model-comparison}
  \centering
  \begin{tabular}{l p{4.5 cm} l r c}
    \toprule
    \textbf{Model} & \textbf{Latent \& Decoding Approach} & \textbf{Regularization} & \textbf{Weights} & \textbf{Reduction(\%)} \\
    \midrule
    A & Standard 32 Latent Space \newline Base Decode – Dense & KL Divergence & 896,198 & 0.00\% \\
    B & Standard 32 Latent Space \newline Base Decode – Dense & Sliced Wasserstein & 896,198 & 0.00\% \\
    C & Dynamic Latent from Sliding Win (160\,ms, Stride=40\,ms) \newline Base Decode – Dense & Sliced Wasserstein & 491,656 & 45.13\% $\downarrow$ \\
    D & Dynamic Latent from Sliding Win (160\,ms, Stride=40\,ms) \newline Base Decode – de-convolution & Sliced Wasserstein & 487,624 & 45.58\% $\downarrow$ \\
    \bottomrule
  \end{tabular}
\end{table}
\begin{table}[ht]
  \caption{Parameters (w.r.t. target  band) used in Sub-Window Convolution Encoding Block}
  \label{tab:eeg-processing}
  \centering
  \scriptsize
  \begin{tabular}{c c c c c c c c c c c}
    \toprule
    \textbf{Iter.} & \textbf{Target Band} & \textbf{Filter} & \textbf{Inp Win} & \textbf{$F_S$} & \textbf{Kernel} & \textbf{$f_c$} & \textbf{Sub-Win} & \textbf{$f_L$} & \textbf{Stride} & \textbf{Out Win} \\
    & \textbf{(Hz)} & & \textbf{$W^T_i$} & \textbf{(Hz)} & \textbf{$K$} & \textbf{($F_S/K$)} & \textbf{$W_S$} & \textbf{($F_S/W_S$)} & & \textbf{$W^T_{i+1}$} \\
    \midrule
    1 & Beta (12–30)  & 16  & 256 & 100 & 5  & 20.00 Hz & 8   & 12.50 Hz & 1 & 256 \\
    2 & Alpha (8–12)  & 32  & 256 & 100 & 9  & 11.11 Hz & 16  & 6.25 Hz  & 2 & 128 \\
    3 & Theta (4–8)   & 64  & 128 & 50  & 9  & 5.55 Hz  & 16  & 3.12 Hz  & 2 & 64  \\
    4 & Delta (0.5–4) & 128 & 64  & 25  & 7  & 3.57 Hz  & 64  & 0.39 Hz  & 1 & 64  \\
    \bottomrule
  \end{tabular}
\end{table}
\begin{table}[ht]
  \caption{Parameters (w.r.t. target  band) used in Sub-Window Convolution Decoding Block}
  \label{tab:eeg-filtering}
  \centering
  \scriptsize
  \begin{tabular}{c c c c c c c c c c c}
    \toprule
    \textbf{Iter.} & \textbf{Target Band} & \textbf{Filter} & \textbf{Inp Win} & \textbf{Stride} & \textbf{Out Win} & \textbf{$F_S$} & \textbf{Kernel} & \textbf{$f_c$} & \textbf{Sub-Win} & \textbf{$f_L$} \\
    & \textbf{(Hz)} & & \textbf{$W^T_{i-1}$} & & \textbf{$W^T_i$} & \textbf{(Hz)} & \textbf{$K$} & \textbf{($F_S/K$)} & \textbf{$W_S$} & \textbf{($F_S/W_S$)} \\
    \midrule
    1 & Delta (0.5–4.0)  & 64  & 64  & 0.5 & 128 & 50  & 13 & 3.84 Hz  & 128 & 0.39 Hz  \\
    2 & Theta (4.0–8.0)  & 32  & 128 & 0.5 & 256 & 100 & 15 & 6.66 Hz  & 32  & 3.12 Hz  \\
    3 & Alpha (8.0–12.0) & 16  & 256 & 1   & 256 & 100 & 9  & 11.11 Hz & 16  & 6.25 Hz  \\
    4 & Beta (12–30)     & 1   & 256 & 1   & 256 & 100 & 5  & 20.00 Hz & 8   & 12.50 Hz \\
    \bottomrule
  \end{tabular}
\end{table}

\subsection{Additional Figures}
\begin{figure}[H]
    \centering
    \includegraphics[width=\linewidth]{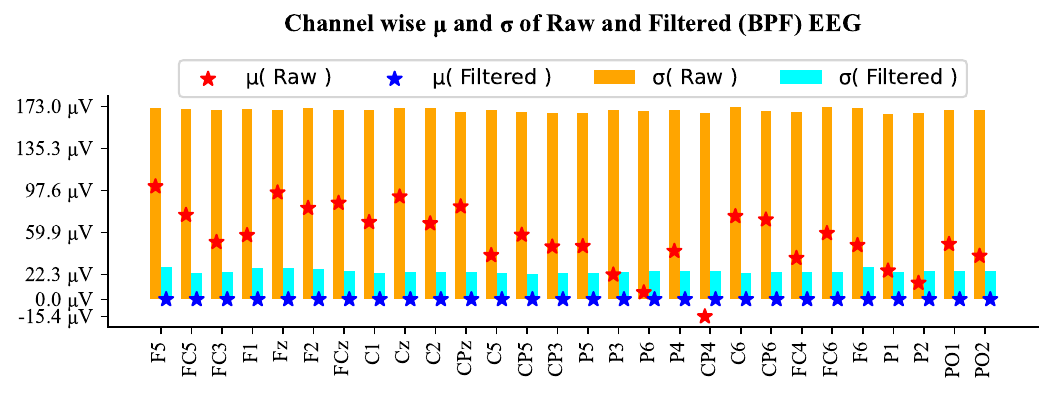}
    \caption{Channel-wise mean and standard deviation of EEG signals for subject `a` before and after band-pass filtering (BPF).}
    \label{fig:Fig5}
\end{figure}

\begin{figure}[t]
    \centering
    \includegraphics[width=\linewidth]{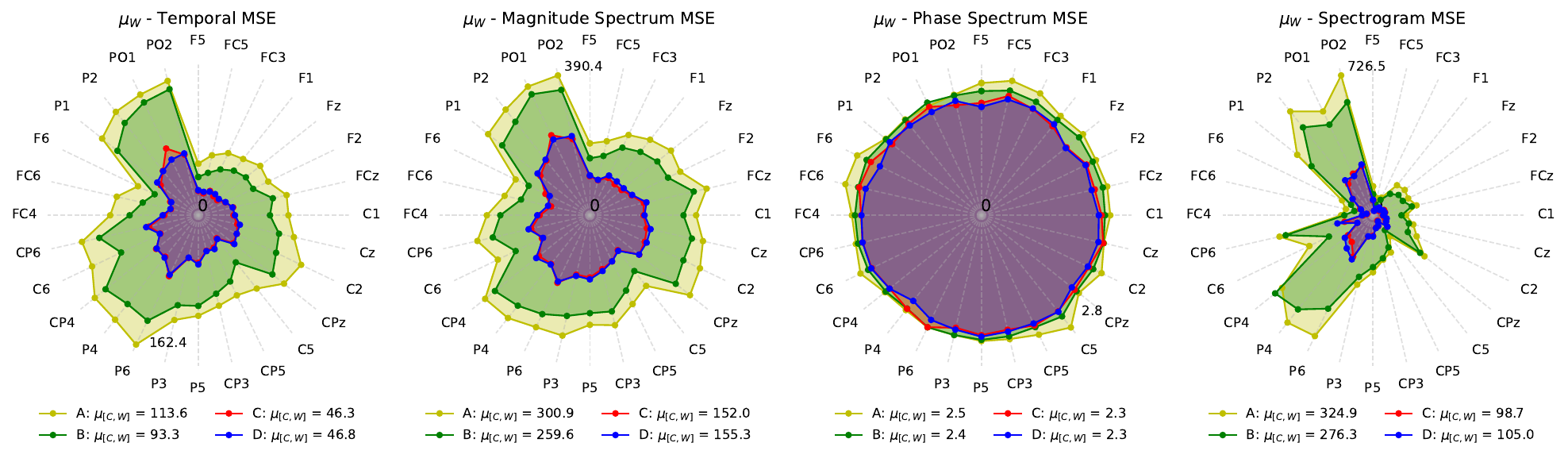}
    \caption{Comparison of reconstruction error across models while reconstructing clean EEG segments ($\leq\pm3.5 \sigma$) of subject ‘a’ using four MSE-based metrics. Radar plots illustrate the performance of Models A–D across EEG channels for (a) Temporal MSE, (b) Magnitude Spectrum MSE, (c) Phase Spectrum MSE, and (d) Spectrogram MSE.}
    \label{fig:Fig6}
\end{figure}
\begin{figure}[H]
    \centering
    \includegraphics[width=\linewidth]{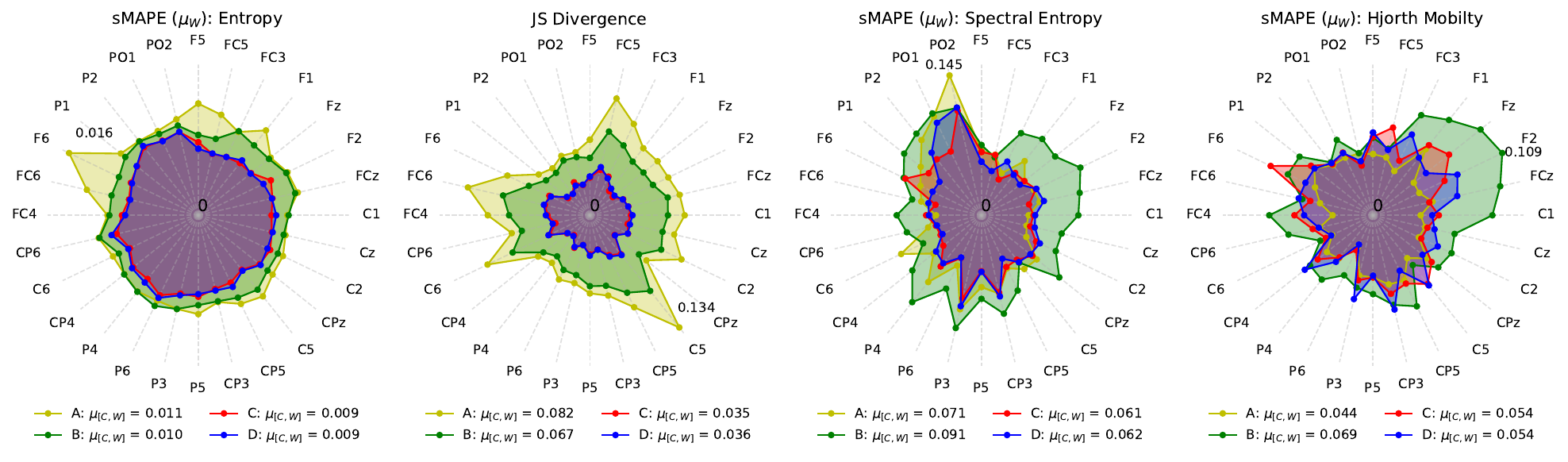}
    \caption{Comparison of model performance across EEG channels using entropy-based and distributional metrics. Radar plots show the reconstruction accuracy of Models A–D for (a) Entropy, (b) JS Divergence, (c) Spectral Entropy, and (d) Hjorth Mobility, evaluated on clean EEG segments ($\leq\pm3.5\sigma$)  of Subject ‘a’. Metrics are reported as symmetric Mean Absolute Percentage Error (sMAPE) or divergence values, with lower values indicating better performance.}
    \label{fig:Fig7}
\end{figure}
\begin{figure}[b]
    \centering
    \includegraphics[height=5 cm]{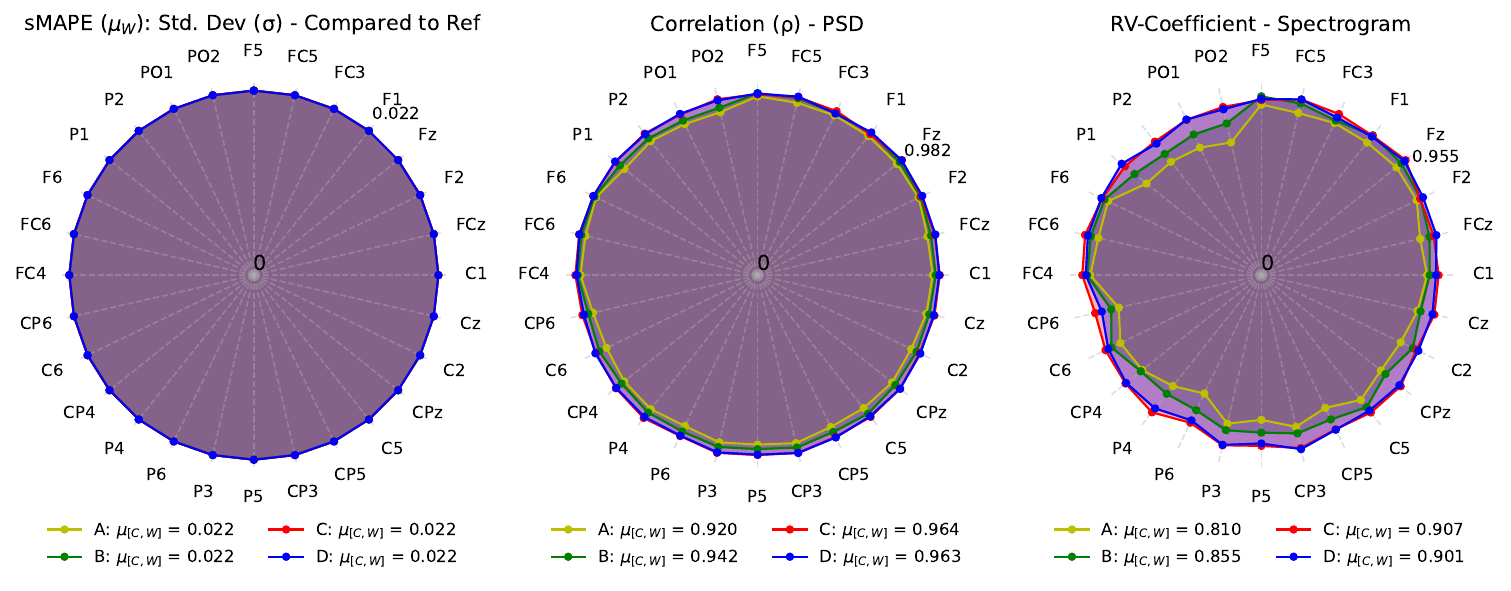}
    \caption{Comparison of reconstruction quality across models for noisy EEG signals ($>\pm3.5\sigma$) from Subject ‘a’ data. Signals were scaled using reference statistics ($\mu_{Ref},\sigma_{Ref}$) derived from preceding clean segments. Radar plots show model-wise performance across EEG channels for: (a) sMAPE of standard deviation relative to reference ($\sigma_{Ref}$) (b) Pearson correlation ($\rho$) of power spectral density (PSD), and (c) RV-Coefficient of the spectrogram.. All models show consistent scaling, confirming the effectiveness of the \texttt{ScaleOutput} layer.}
    \label{fig:Fig8}
\end{figure}
\begin{figure}[H]
    \centering
    \includegraphics[width=\linewidth]{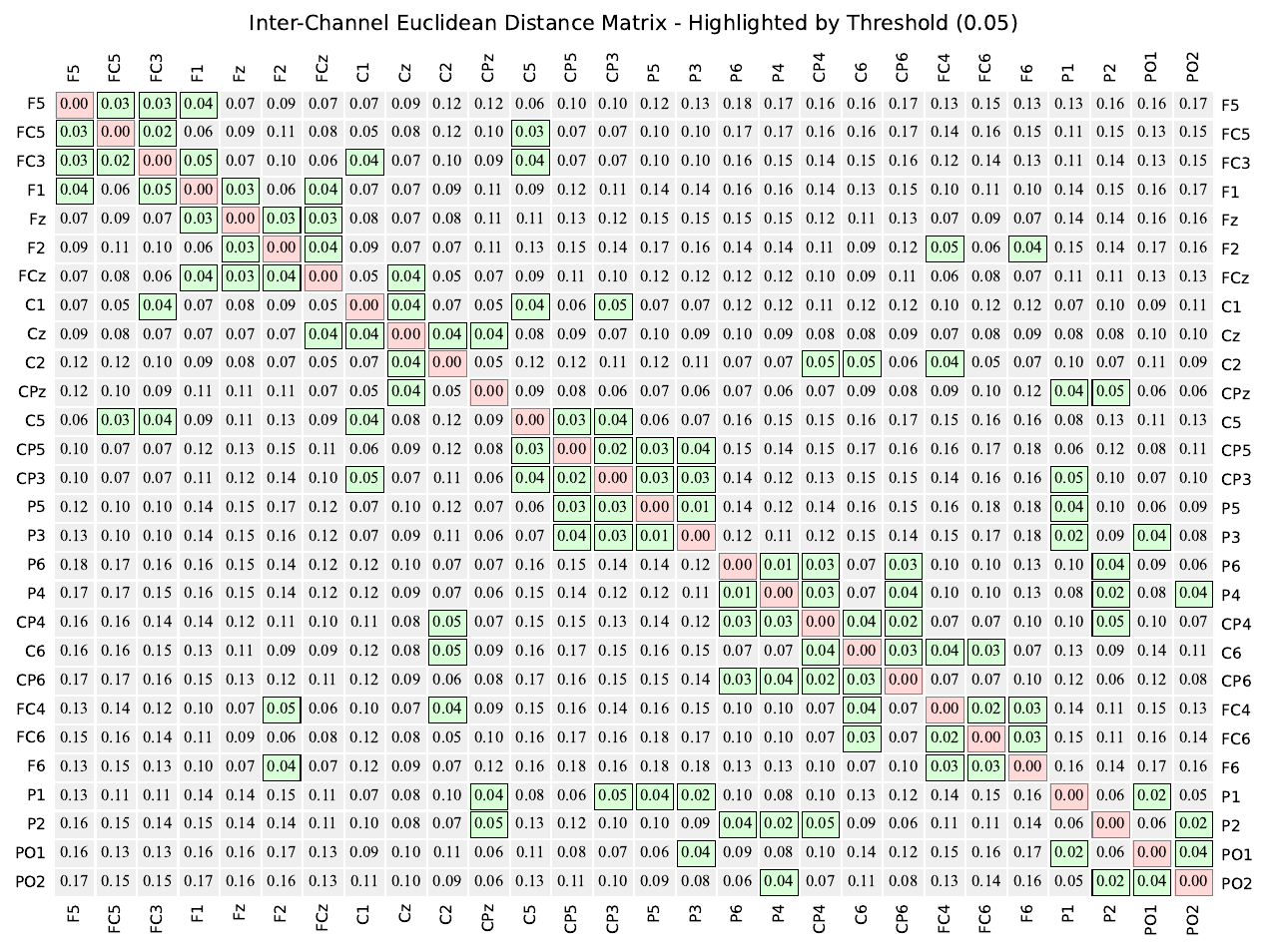}
    \caption{Inter-channel Euclidean distance matrix based on the 10–20 EEG electrode placement standard. Each row represents a reference EEG channel (red cell), and green cells indicate neighboring channels whose Euclidean distance from the reference is below or equal to the threshold of 0.05. This spatial neighborhood mapping was used to define channel-specific subspaces.}
    \label{fig:Fig9}
\end{figure}

\begin{figure}[H]
    \centering
    \includegraphics[width=\linewidth]{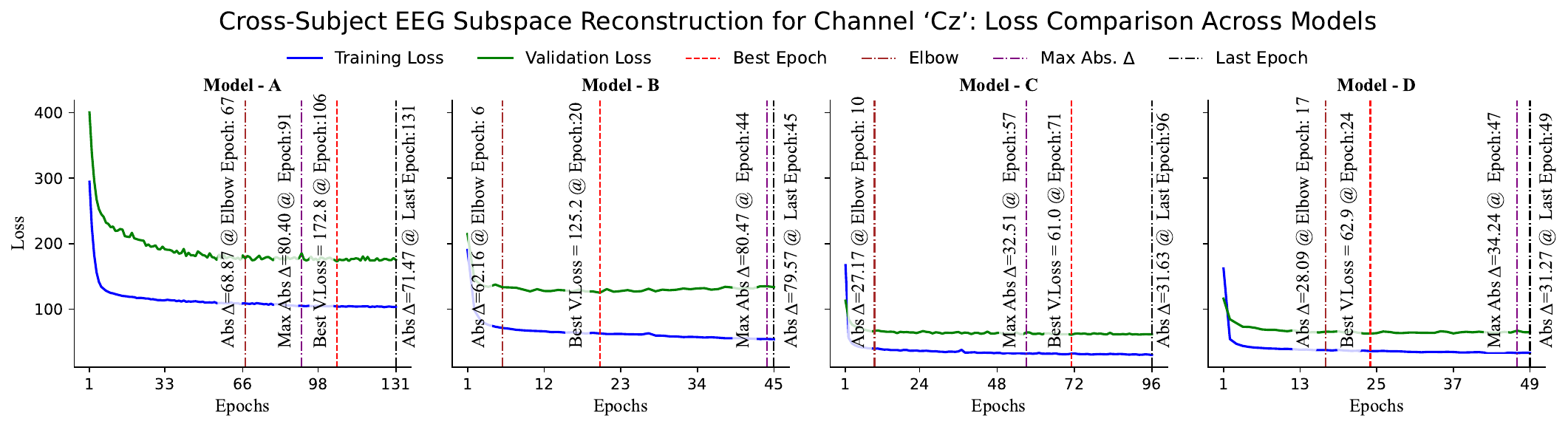}
    \caption{Training and validation loss curves for four models (A–D), trained using clean EEG windows from subjects ‘b’, ‘c’, and ‘d’ to reconstruct the EEG subspace of Channel Cz for subject ‘a’. The blue line represents training loss, while the green line denotes validation loss. Key metrics—including the Elbow Point, Best Validation Loss, Maximum Absolute Difference ($\Delta$), and Final Epoch $\Delta$—are annotated for each model. Model A exhibits prolonged training with poor generalization and pronounced overfitting. Model B achieves improved validation loss but still shows signs of overfitting. Models C and D demonstrate superior generalization, characterized by lower validation loss and minimal overfitting.}
    \label{fig:Fig10}
\end{figure}

\begin{figure}[H]
    \centering
    \includegraphics[width=\linewidth]{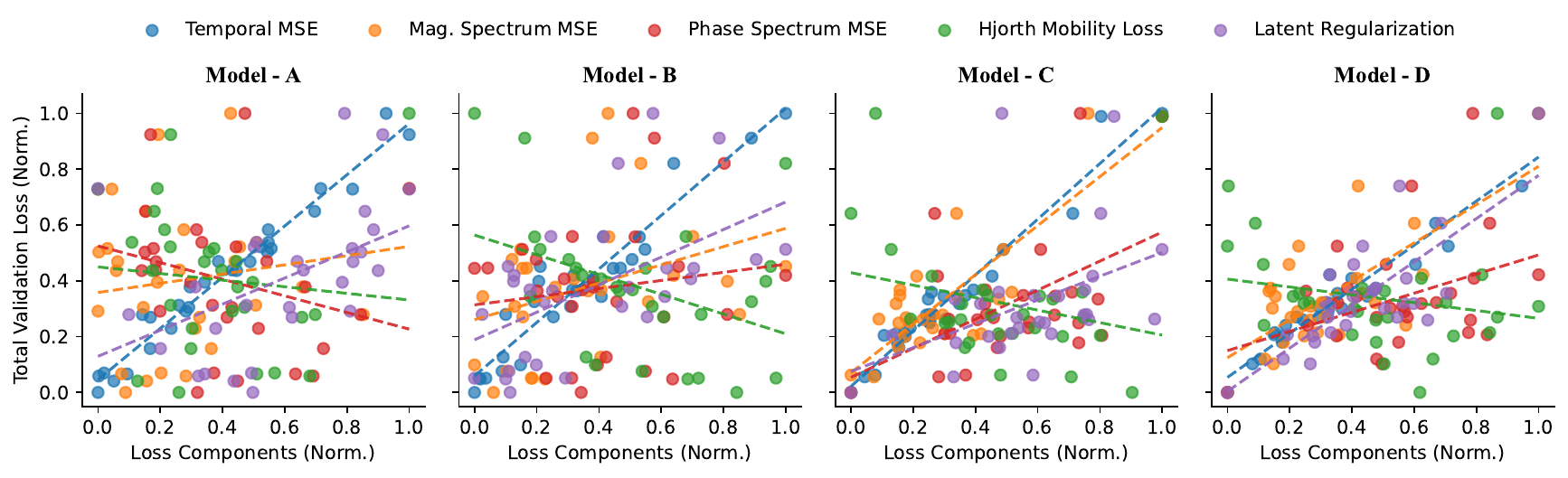}
    \caption{Scatter plots showing the correlation between Total Validation Loss (y-axis) and individual validation loss components (x-axis) across Models A–D at the best epoch, evaluated over 28 channels. Each color represents a different loss component: Temporal MSE (blue), Magnitude Spectrum MSE (orange), Phase Spectrum MSE (red), Hjorth Mobility (green), and Latent Regularization (purple; KLD in Model A, SWD in Models B–D). Trend lines indicate the strength and direction of correlation.}
    \label{fig:Fig11}
\end{figure}
\begin{figure}[H]
    \centering
    \includegraphics[width=0.9\linewidth]{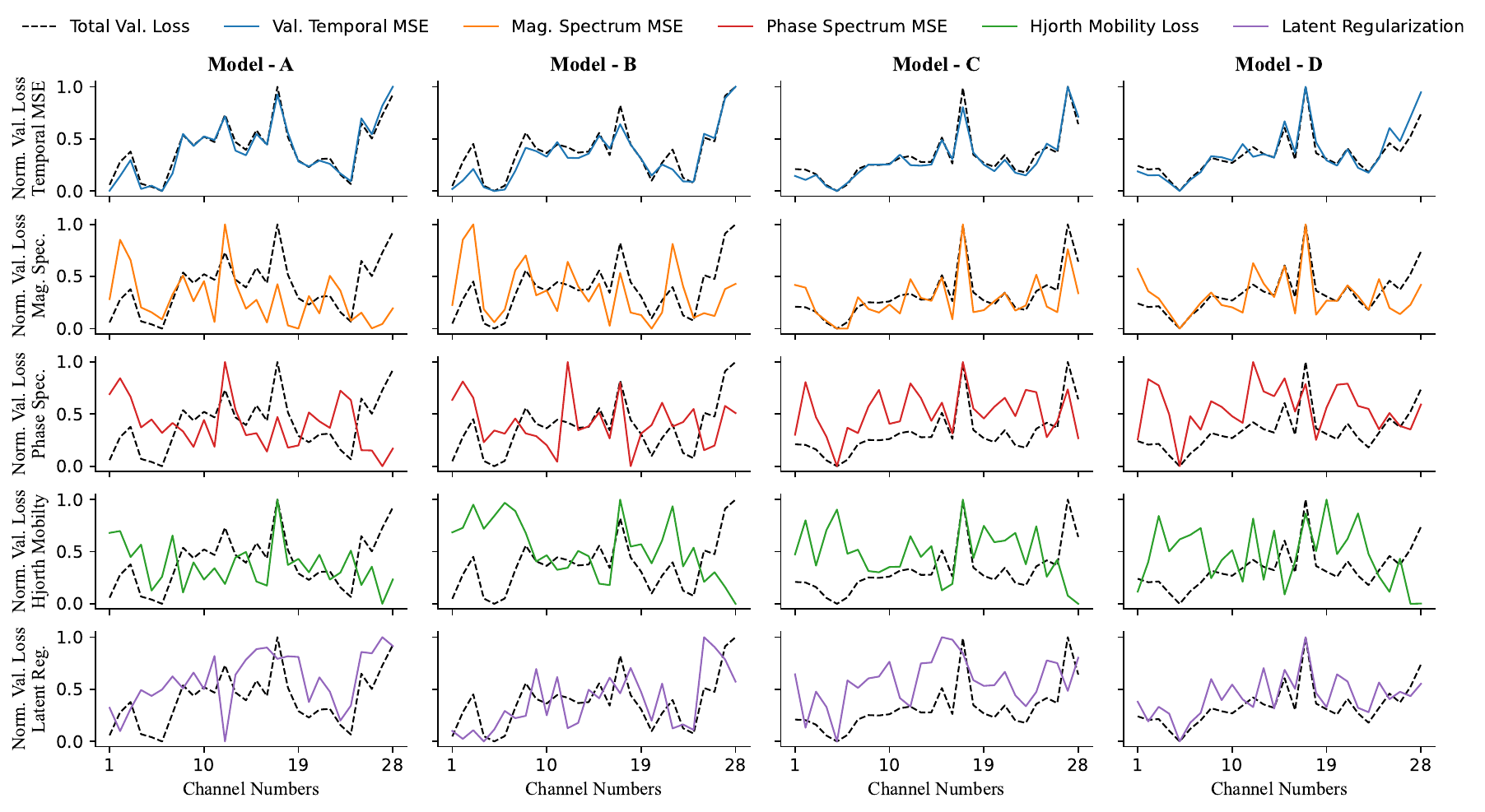}
    \caption{Subplots showing channel-wise comparison of individual validation loss components (Temporal MSE, Magnitude Spectrum MSE, Phase Spectrum MSE, Hjorth Mobility, Latent Regularization-KLD/SWD) against Total Validation Loss (dashed black line) across Models A–D at the best epoch. Each row corresponds to a specific loss component, and each column represents a model. Models were trained independently per channel using clean data from subjects 'b', 'c', and 'd' (tested on subject 'a'). The layout highlights how each component tracks with total loss across channels.}
    \label{fig:Fig12}
\end{figure}
\begin{figure}[H]
    \centering
    \includegraphics[width=0.55\linewidth]{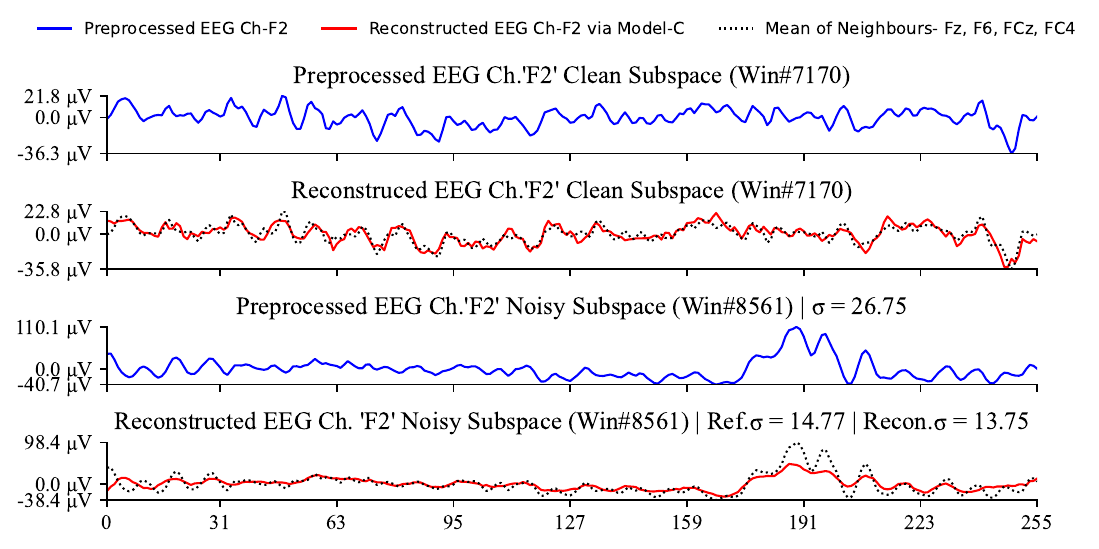}
    \caption{Comparison of clean and noisy EEG signals from channel F2 (subject`a') with Model-C reconstructions, highlighting alignment with neighboring channels and signal variability.}
    \label{fig:Fig13}
\end{figure}
\begin{figure}[H]
    \centering
    \includegraphics[width=\linewidth]{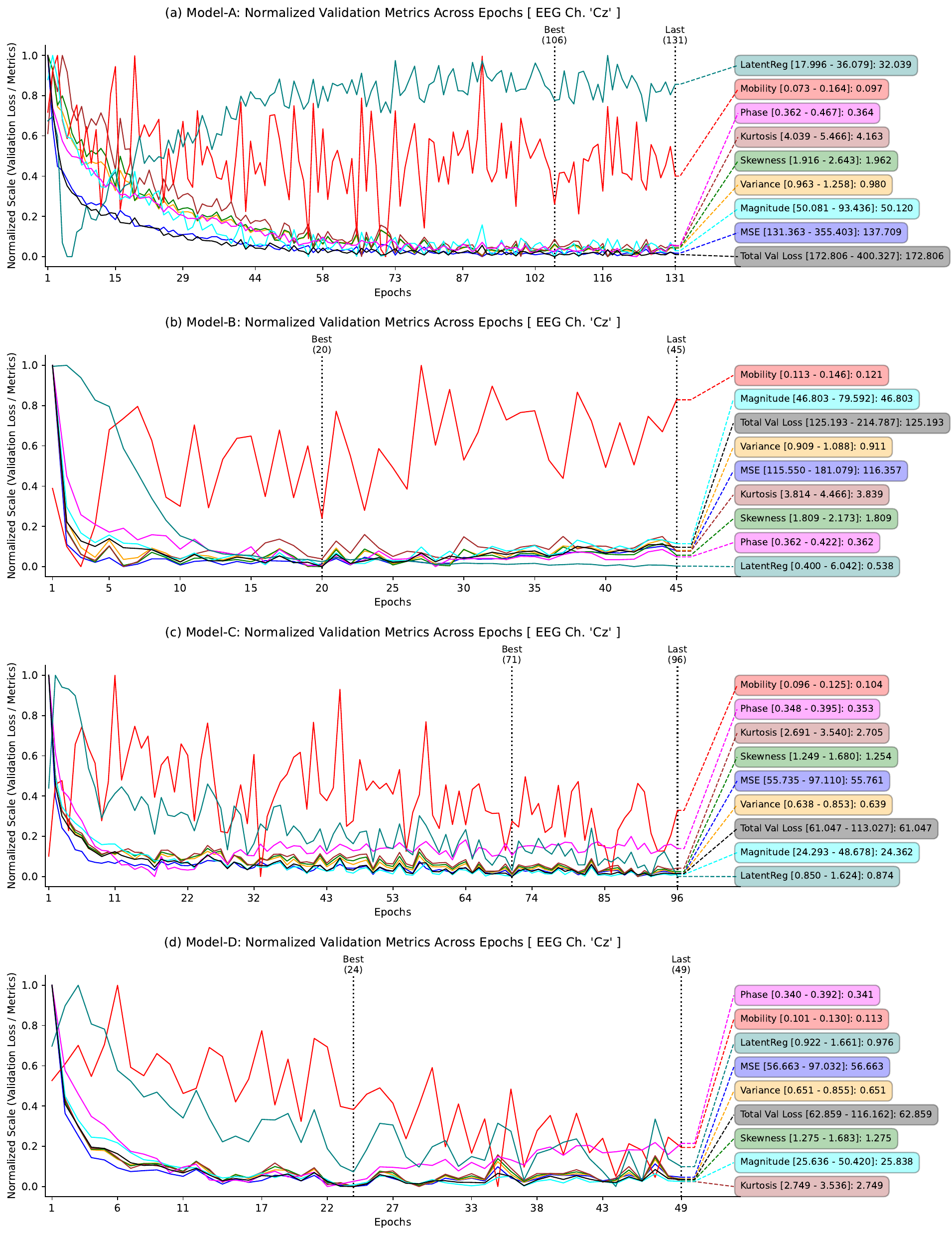}
    \caption{Normalized validation metrics across training epochs for Models A–D, trained on clean EEG data from subjects ‘b’, ‘c’, and ‘d’ (tested on subject ‘a’) for Cz-channel reconstruction, reveal distinct convergence behaviors. Each model’s Total Validation Loss reflects contributions from Temporal MSE, Magnitude/Phase Spectrum MSE, Mobility, and Latent Regularization (SWD), alongside auxiliary metrics—Kurtosis, Skewness, and Variance—normalized to [0, 1] for comparability. Model-A demonstrates early high performance but exhibits pronounced fluctuations beyond the initial epochs, suggesting potential overfitting and reduced generalization. In contrast, Model-B and Model-D converge more rapidly and maintain stable validation trajectories, indicating better optimization dynamics and robustness. The ‘Best’ epoch was identified as the point of minimum total validation loss, while training termination was governed by a patience setting of 25 epochs.}
    \label{fig:Fig14}
\end{figure}
\begin{figure}[t]
    \centering
    \includegraphics[width=\linewidth]{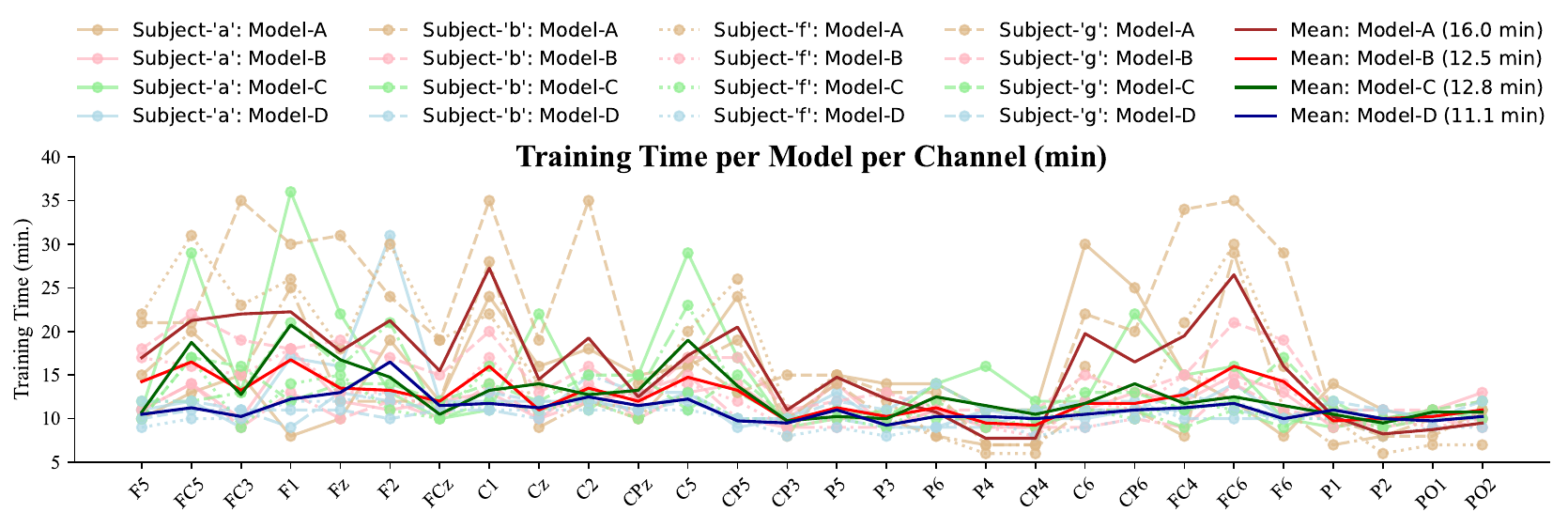}
    \caption{Training time per model across EEG channels and subjects. Mean training durations are highlighted for each model: Model-A (16 min), Model-B (12.5 min), Model-C (12.8 min), and Model-D (11.1 min), illustrating comparative computational efficiency.}
    \label{fig:Fig15}
\end{figure}
\begin{figure}[H]
    \centering
    \includegraphics[width=\linewidth]{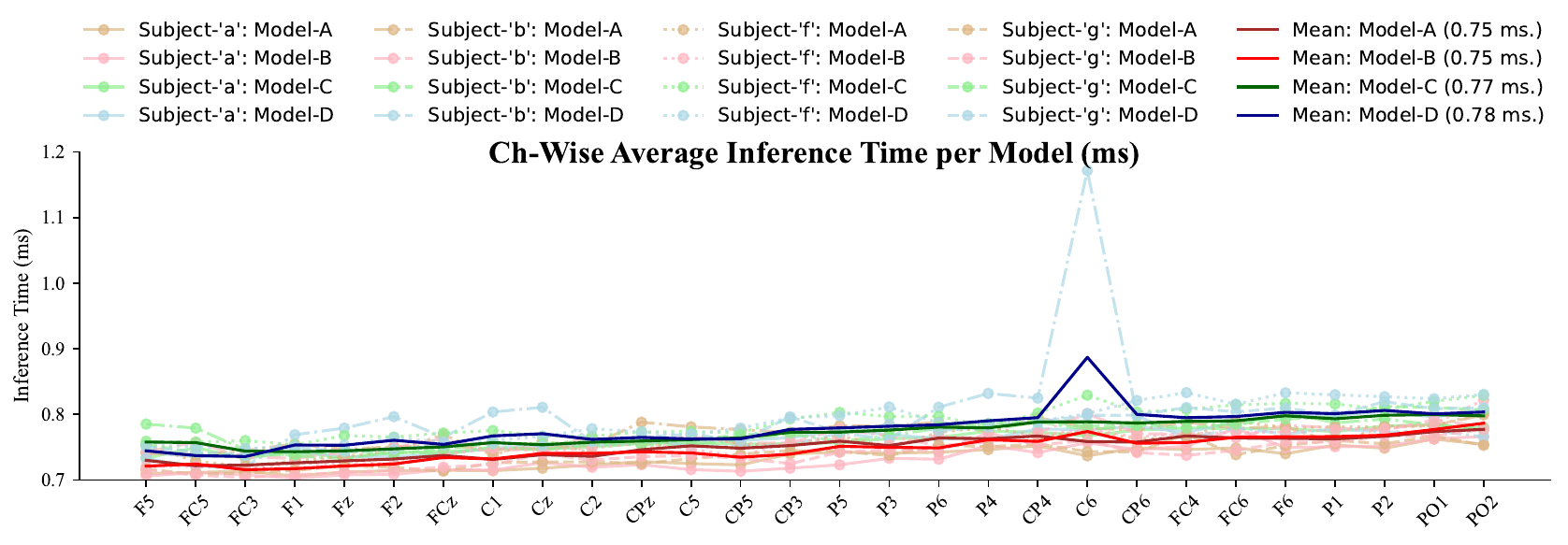}
    \caption{Channel-wise average inference time per model across subjects. Solid lines indicate mean inference times: Model-A (0.75 ms), Model-B (0.77 ms), Model-C (0.78 ms), and Model-D (0.78 ms). All models demonstrate sub-millisecond latency with minimal channel-wise variation, supporting suitability for real-time EEG applications.}
    \label{fig:Fig16}
\end{figure}
\end{document}